\documentclass[sigconf,nonacm]{acmart} 

\usepackage{balance}

\usepackage[most]{tcolorbox}  

\usepackage{listings}
\usepackage{xcolor}  

\usepackage{afterpage}
\usepackage{algorithmic}
\usepackage{array}
\usepackage{booktabs}
\usepackage{caption}
\usepackage[compress]{cite}
\usepackage{cleveref}
\usepackage{enumitem}
\usepackage{float}
\usepackage{fontawesome5}
\usepackage[T1]{fontenc}
\usepackage{graphicx}
\usepackage{hyperref}
\usepackage{lipsum}
\usepackage{lscape}
\usepackage{makecell}
\usepackage{multirow}
\usepackage{siunitx}
\usepackage{soul}
\usepackage{natbib}[numbers]
\usepackage{tabularx}
\usepackage{tcolorbox}
\usepackage{textcomp}
\usepackage{titlesec}
\usepackage[normalem]{ulem}
\usepackage{xspace}
\usepackage{balance}

\titlespacing*{\section}
{0pt}{6pt}{4pt}

\titlespacing*{\subsection}
{0pt}{5pt}{3pt}

\titlespacing*{\subsubsection}
{0pt}{4pt}{2pt}

\newif\ifDEBUG
\DEBUGfalse

\newcommand{\myparagraph}[1]{\vspace{0.20cm}\noindent\textbf{#1} \noindent{}}
\newcommand{\cveparagraph}[1]{\vspace{0.20cm}\noindent\underline{#1} \noindent{}}

\ifDEBUG
    \newcommand{\samiha}[1]{{\small{\textcolor{purple}{\textit{\textbf{Samiha}: #1}}} }}
    \newcommand{\NMS}[1]{{\small{\textcolor{red}{\textit{\textbf{Nicholas}: #1}}} }}
    \newcommand{\mona}[1]{{\small{\textcolor{blue}{\textit{\textbf{MR}: #1}}} }}
    \newcommand{\george}[1]{{\small{\textcolor{orange}{\textit{\textbf{gkt}: #1}}} }}
    \newcommand{\TODO}[1]{\hl{TODO: #1}}
\else
    \newcommand{\samiha}[1]{}
    \newcommand{\NMS}[1]{}
    \newcommand{\mona}[1]{}
    \newcommand{\george}[1]{}
    \newcommand{\TODO}[1]{}
\fi

\AtBeginDocument{
  }

\setcopyright{acmlicensed}
\copyrightyear{2026}
\acmYear{2026}
\acmDOI{XXXXXXX.XXXXXXX}
\acmConference[Conference acronym 'XX]{Make sure to enter the correct
  conference title from your rights confirmation email}{June 03--05,
  2018}{Woodstock, NY}

\acmISBN{978-1-4503-XXXX-X/2018/06}

\begin{document}

\title{Process-based Indicators of Vulnerability Re-Introducing\\ Code Changes: An Exploratory Case Study}

\author{Samiha Shimmi}
\authornote{Samiha Shimmi and Nicholas M. Synovic contributed equally to this work (co-first authors).}
\affiliation{
  \institution{Northern Illinois University}
  \city{Dekalb, IL}
  \country{USA}}
  \orcid{0000-0002-5307-8819}
\email{sshimmi@niu.edu}

\author{Nicholas M. Synovic}
\authornotemark[1]
\affiliation{
  \institution{Loyola University Chicago}
  \city{Chicago, IL}
  \country{USA}}
  \orcid{0000-0003-0413-4594}
\email{nsynovic@luc.edu}

\author{Mona Rahimi}
\authornote{Mona Rahimi and George K. Thiruvathukal contributed equally to this work (co-supervisors).}
\affiliation{
  \institution{Northern Illinois University}
  \city{Dekalb, IL}
  \country{USA}}
  \orcid{0000-0001-7228-7520}
\email{rahimi@cs.niu.edu}

\author{George K.~Thiruvathukal}
\authornotemark[2]
\affiliation{
  \institution{Loyola University Chicago}
  \city{Chicago, IL}
  \country{USA}}
  \orcid{0000-0002-0452-5571}
\email{gthiruvathukal@luc.edu}

\renewcommand{\shortauthors}{Shimmi, Synovic, Rahimi, and Thiruvathukal}

\begin{abstract}
Software vulnerabilities often persist or re-emerge after being fixed, revealing the complex interplay between code evolution and socio-technical factors.
While source code metrics provide useful indicators of vulnerabilities, software engineering process metrics can measure engineering activities that lead to their introduction. 
To our knowledge, few studies have explored whether these metrics can identify these activities over time --- insights that are essential for anticipating and mitigating software vulnerabilities. 
This work highlights the critical role of process metrics along with code changes in understanding and mitigating \textit{vulnerability reintroduction}. 
We move beyond file-level prediction and instead analyze vulnerability fixes at the commit level, focusing on sequences of changes through which vulnerabilities re-emerge.

Our approach emphasizes that reintroduction is rarely the result of one isolated action, but emerges from cumulative engineering activities and socio-technical conditions.
To support this analysis, we conduct a case study on the ImageMagick project by correlating longitudinal process metrics including bus factor, issue density, and issue spoilage with vulnerability reintroduction activities, encompassing 76 instances of reintroduced vulnerabilities.
Our findings show that vulnerability reintroduction often aligns with increased issue spoilage and fluctuating issue density, reflecting short-term inefficiencies in issue management and team responsiveness.
These observations provide a foundation for broader studies that combine process and code metrics to predict vulnerability reintroducing fixes and strengthen software security.
\end{abstract}



\maketitle
\begin{figure}[h]
    \centering
    \setlength{\abovecaptionskip}{2pt}
    \setlength{\belowcaptionskip}{0pt}
    \includegraphics[width=0.99\linewidth]{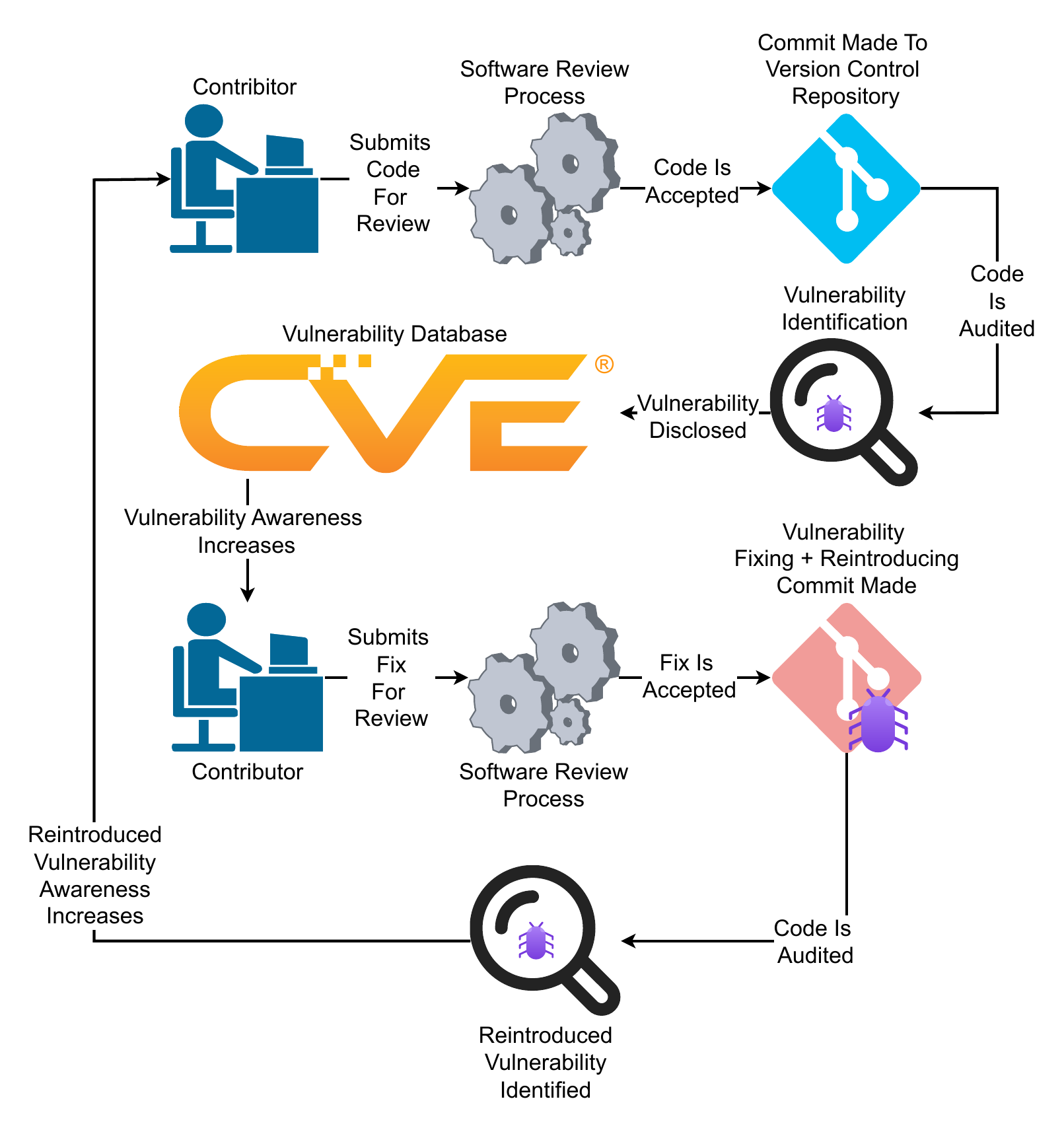}
    \caption{
    Overview of the vulnerability reintroduction process.
    Contributors submit code that passes review and enters the version control system. 
    Once a vulnerability is disclosed and patched, a new flaw is inadvertently reintroduced in the fix, leading to a recurring cycle of identification and patching.
    }
    \label{fig:figF}
\end{figure}

\section{Introduction}
Software security remains a critical challenge within open-source software. 
The rapid pace and complexity of technological advancement place software engineers under time constraints, increasing the risk of flawed implementations and introducing vulnerabilities.
Software vulnerabilities are reported to the 
    National Vulnerability Database (NVD)~\citep{national_institute_of_standards_and_technology_national_2025} and
    Common Vulnerabilities and Exposures (CVE)~\citep{the_mitre_corporation_common_2025}
databases.

Despite software engineers' efforts to address vulnerabilities, their fixes can inadvertently introduce new security issues~\citep{sliwerski2005changes, shimmi_mining_2022} as visualized in~\Cref{fig:figF}. 
Vulnerable software systems 
    compromise overall system dependability~\citep{braz_exploratory_2022}, 
    hamper reproducibility~\citep{kaur_analysis_2021}, and 
    diminish trust~\citep{wermke_committed_2022} 
    
Moreover, the integration of software dependencies enables vulnerabilities to propagate through software supply chains~\citep{zimmermann_small_2019, zahan_what_2022, hegewald_evaluating_2025}.
Although automated 
    security audits~\citep{zahan_openssf_2023, snyk_limited_snyk_2025, google_open_2025, github_about_2025} and 
    tooling~\citep{nong_open_2023, alazmi_systematic_2022, nie_mvdetecter_2022, sheng_llms_2025}
are available to the open-source community, the impact of these tools on engineering decisions is limited~\citep{eisty_survey_2018, ayala_mixed-methods_2025}. 
While research has examined software vulnerability prediction~\citep{nong_open_2023, sheng_llms_2025}, to our knowledge, a gap persists in identifying vulnerability reintroducing activities where remediation of one vulnerability inadvertently introduces another~\citep{shimmi_mining_2022}.
Furthermore, as vulnerabilities propagate through commit operations, a gap exists in analyzing the software engineering process involved.

To address this, we conduct the first exploratory case study of a software's engineering process and its correlation to vulnerability reintroduction.
We identify pairs of vulnerability reintroducing and fixing commits and automatically compute longitudinal process metrics for 
    bus factor, 
    issue density, and 
    issue spoilage 
for the ImageMagick~\citep{ImageMagick} project --- a well known image processing project.
Our results show that 
    (1) ImageMagick has maintained a healthy number of contributors, implying that there is always a contributor available to fix and review vulnerability reintroducing commits;
    (2) ImageMagick’s issue density has stayed low even during vulnerability reintroductions, demonstrating that maintainers can manage incoming issues efficiently while developing new features, suggesting the project’s workflow can handle backlog accumulation without disrupting regular development activities; and
    (3) mageMagick’s issue backlog spikes during vulnerability reintroductions, suggesting that contributors prioritize remediation over routine issue closure. This shift in focus toward security fixes creates a bottleneck in parallel task management, straining resources during periods of high-security workloads.

Taken together, these findings suggest that while ImageMagick’s contributor base and issue management processes provide a strong foundation for sustainable maintenance, vulnerability reintroductions still impose measurable strain on project dynamics. 
This underscores the need for improved process-aware tooling and decision support systems that help engineers anticipate, detect, and mitigate reintroduction risks before they propagate through the codebase.

Our contributions are:
    (1)The first empirical case study correlating longitudinal software engineering process metrics with open-source software engineering vulnerability reintroduction activities;
    (2) A methodology for identifying vulnerability reintroducing commits leveraging existing algorithms~\citep{sliwerski_when_2005} and LLMs; and
    (3) A dataset of 76 vulnerability-reintroducing commits that extends existing vulnerability datasets by including future commits that fix the reintroduced vulnerabilities.

We release our source code and dataset artifacts on Zenodo~\citep{shimmi_2026_18251736}.

\section{Background and Related Work}
In this section, we give a brief overview of software engineering process metrics, process metrics for predicting software vulnerabilities, and vulnerability reintroduction.

\subsection{Software Engineering Process Metrics}
Software engineering process metrics quantify the activity involved in the creation or maintenance of a software system~\citep{eisty_survey_2018}.
Prior work has shown that 
    code complexity, 
    churn, and 
    developer activity metrics 
can serve as indicators of software vulnerabilities~\citep{shin2010evaluating}, however the open-source community has seen limited adoption of these~\citep{eisty_survey_2018}.
These metrics are computed at specific releases, rather than examining longitudinal trends.
Captured at a specific release, snapshot process metrics offer insights into the software system's current state, however, measuring process metrics longitudinally captures socio-technical trends influencing project evolution~\citep{synovic_snapshot_2022}.
Several process metrics have been proposed~\citep{fritz_degree--knowledge_2014, goggins_open_2021, linux_foundation_community_nodate}, but for our study, we focus on derived process metrics including
    bus factor,
    issue density, and
    issue spoilage.

\underline{Bus Factor} measures team-wide familiarity with a system's components by counting contributions to various elements~\citep{cosentino_assessing_2015}; high bus factors indicate widespread understanding, while a low bus factor (i.e., less than one~\citep{piggot_how_2013}) suggests siloed knowledge potentially limiting the number of active contributors who could resolve vulnerabilities.
\underline{Issue Density} extends the defect density metric~\citep{fenton_software_2014} to open-source issue trackers; high density suggests engineers struggle to manage community submissions efficiently, whereas low density implies timely responses to these inputs.
\underline{Issue spoilage} quantifies how quickly engineering teams close reported issues; high spoilage indicates unaddressed lingering issues, whereas low spoilage signifies prompt handling of issues.
A detailed overview of each of these metrics is presented in~\citep{thiruvathukal_metrics_2018, synovic_snapshot_2022}.

Our study advances software engineering process metrics by analyzing several process metrics longitudinally to correlate engineering activities to vulnerability reintroduction. 
This approach underscores the critical gap between awareness and practical adoption of such metrics in software engineering communities, emphasizing their potential as actionable indicators for improving code quality and security resilience.

\subsection{Software Vulnerabilities and Vulnerability Reintroduction}
A software vulnerability refers to specific weakness in a software system's code that can be exploited to compromise the 
    confidentiality, 
    integrity, or 
    availability 
of a system. 
Centralized databases 
    aggregate, 
    report, 
    score, and 
    distribute 
publicly disclosed vulnerabilities to enable vulnerability tracking and awareness~\citep{national_institute_of_standards_and_technology_national_2025, the_mitre_corporation_common_2025, google_open_2025, github_github_2025, gitlab_gitlab_2025}.
Vulnerabilities can be classified via the Common Weakness Enumeration (CWE) schema~\citep{the_mitre_corporation_common_2025} which captures recurring weaknesses in software (e.g., buffer overflows, improper input validation) that automated tooling can classify~\citep{das_v2w-bert_2021, han_learning_2017}.
Furthermore, the Common Vulnerability Scoring System (CVSS) provides a means to quantify the severity of a vulnerability from a range of zero to ten~\citep{forum_of_incident_response_and_security_teams_common_2024}, with ten being reserved for the most severe vulnerabilities~\citep{benny_isaacs_redis_2025,chen_zhaojun_apache_2021}.

Software fixes attempt to address and remove software vulnerabilities, but prior work has found that not all fixes are permanent~\citep{sliwerski2005changes, braz_exploratory_2022}.
Vulnerability reintroduction occurs when a fix inadvertently reintroduces a new security weakness as confirmed by later fixes~\citep{shimmi_mining_2022,braz_exploratory_2022}.
Software vulnerability reintroduction was conceptually identified as a threat to software systems via \textit{attack–defense co-evolution} by positing that a vulnerability fixing operation might itself serve as a predictor of a subsequent vulnerability~\citep{shimmi2022mining}.
Such reintroductions pose significant risks because they may silently undo prior security improvements and leave systems exposed once again.

While existing datasets have labeled vulnerable and non-vulnerable source code~\citep{chen_diversevul_2023, fan_cc_2020}, to our knowledge, there are no datasets that specifically capture sequences of commits that resolve and reintroduce vulnerabilities.
Our work resolves this by releasing a dataset that extends established vulnerability datasets~\citep{chen_diversevul_2023, fan_cc_2020} with 81 pairs for the ImageMagick project that both resolve existing CWEs while simultaneously reintroducing new vulnerabilities.

\section{Case Study: ImageMagick}
In this section, we introduce our case study on ImageMagick --- a well-known image processing project.
We discuss 
    an overview of the problem, 
    our methodology to evaluate the project's software engineering process metrics and high CVSS scoring CVEs longitudinally, and
    analyze the data collected during our case study.
We selected ImageMagick as our case study due to it being 
    a well-known open-source image processing project with a significant development history~\citep{ImageMagick},
    written primarily in C potentially enabling common weaknesses (e.g., buffer overflows, memory leaks), and
    being readily available in existing vulnerability datasets that have previously mapped CVE disclosures to specific commits~\citep{chen_diversevul_2023,fan_cc_2020}.

\subsection{Problem Overview}
In this section we show an example of a vulnerability being reintroduced as a side effect for resolving a different vulnerability.
This vulnerability remained active for 
    313 days
across 
    1,094 commits
and
    6 releases.

\myparagraph{Vulnerability Reintroducing Fix}
CVE-2018-11625~\citep{the_mitre_corporation_cve-2018-11625_2018} was an out of bounds read vulnerability~\citep{the_mitre_corporation_cwe-125_2025} with a CVSS severity score of 8.8 disclosed on May 31st, 2018. 
Listing 1 shows a commit accepted by ImageMagick that reintroduced a vulnerability while attempting to resolve CVE-2018-11625.
On May 30th, 2018 commit \texttt{5294966} was accepted into ImageMagick to resolve a buffer overflow by allocating an additional byte for \texttt{colormap\_index}.
While the intention was to prevent out-of-bounds writes, this fix was potentially unsafe when \texttt{image->colors} or \texttt{MaxColormapSize} was smaller than \texttt{MaxMap}, creating scenarios where the buffer could still overflow under certain image configurations.\footnote{\url{https://github.com/ImageMagick/ImageMagick/commit/5294966}}

\begin{tcolorbox}[
  title=\textbf{Listing 1: ImageMagick – Initial Fix Commit \texttt{5294966}},
  colback=orange!10, colframe=orange!70!black,
  boxrule=0.5pt, arc=0mm, sharp corners,
  left=3pt, right=3pt, top=2pt, bottom=2pt,
  before skip=4pt, after skip=4pt
]
\begin{lstlisting}[basicstyle=\ttfamily\small, breaklines=true, aboveskip=1pt, belowskip=1pt]
- colormap_index=(unsigned short *) AcquireQuantumMemory(
-   (size_t) image->colors, sizeof(unsigned short));
+ colormap_index=(unsigned short *) AcquireQuantumMemory(
+   (size_t) (image->colors+1), sizeof(unsigned short));
\end{lstlisting}
\label{fig:diffA}
\end{tcolorbox}

\myparagraph{Reintroduced Vulnerability Fix}
Listing 2 shows a subsequent commit \texttt{c111ed9} that corrected the reintroduced vulnerability on April 8th, 2019.
This issue was resolved by ensuring that the allocation size was properly bounded using the \texttt{MagickMax()} function, guaranteeing that the memory allocated for \texttt{colormap\_index} was at least \texttt{MaxMap}.
This change prevented buffer overflow conditions that could occur when the color count was smaller than the maximum mapping size.\footnote{\url{https://github.com/ImageMagick/ImageMagick/commit/280215b9936d145dd5ee91403738ccce1333cab1}}

\begin{tcolorbox}[
  title=\textbf{Listing 2: ImageMagick – Future Fix Commit \texttt{c111ed9}},
  colback=green!5, colframe=green!50!black,
  boxrule=0.5pt, arc=0mm, sharp corners,
  left=3pt, right=3pt, top=2pt, bottom=2pt,
  before skip=4pt, after skip=4pt
]
\begin{lstlisting}[basicstyle=\ttfamily\small, breaklines=true, aboveskip=1pt, belowskip=1pt]
- colormap_index=(unsigned short *) AcquireQuantumMemory(
-   (size_t) (image->colors+1), sizeof(unsigned short));
+ colormap_index=(unsigned short *) AcquireQuantumMemory(
+   MagickMax((size_t) image->colors, MaxMap), sizeof(unsigned short));
\end{lstlisting}
\label{fig:diffB}
\end{tcolorbox}

\begin{figure}[h]
    \centering
    \setlength{\abovecaptionskip}{2pt}
    \setlength{\belowcaptionskip}{0pt}
    \includegraphics[width=\linewidth]{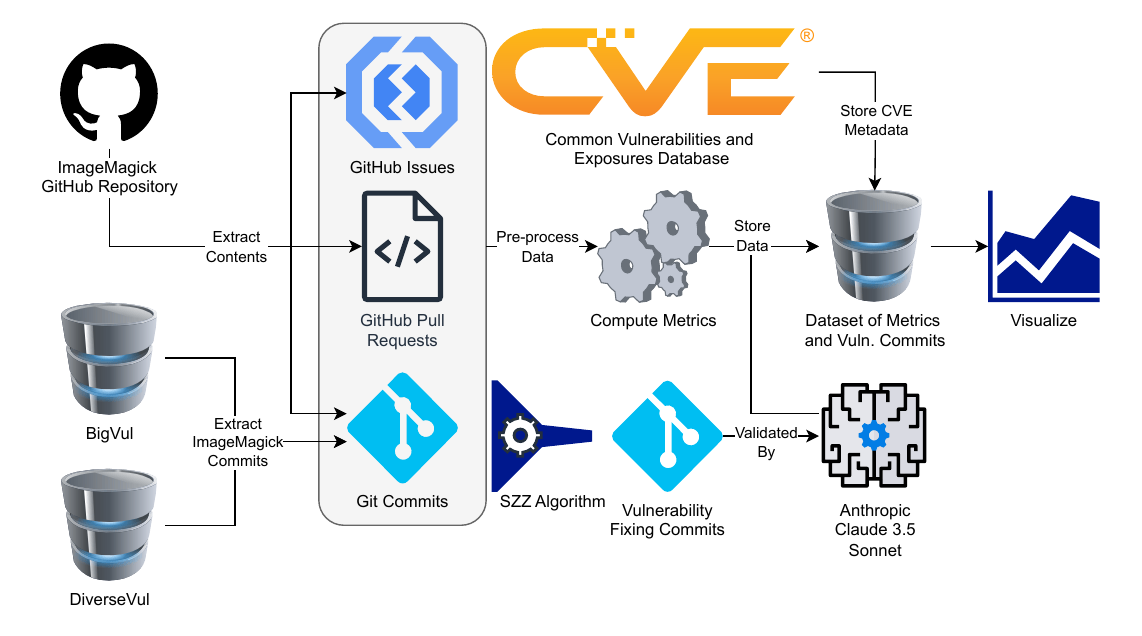}
    \caption{
        We analyzed the ImageMagick project by extracting issues and pull requests from GitHub to calculate issue density and spoilage. 
        Using Git commits from GitHub, BigVul, and DiverseVul, we computed the bus factor and identified vulnerability-fixing commits via the SZZ algorithm and Claude 3.5 Sonnet. 
        Finally, CVE data was integrated to enrich the analysis and facilitate data visualization.
        }
    \label{fig:figJ}
\end{figure}

\subsection{Methodology}
To evaluate the socio-technical context in which vulnerabilities are reintroduced into ImageMagick, we 
    extract and evaluate vulnerability resolving commits from two existing datasets,
    extend existing algorithms and methods with LLMs to identify pairs of reintroducing and fixing commits, and
    leverage prior work on evaluating longitudinal software engineering process metrics on selected high CVSS scoring CVEs that were identified to reintroduce issues.
Our goal is twofold: 
    (1) to flag vulnerability fixing commits that are likely to trigger regressions, and 
    (2) to isolate the engineering process patterns that contribute to vulnerability reintroducing activity in open-source software.
Our methodology is outlined in~\Cref{fig:figJ}

\subsubsection{Dataset Collection}\label{sec:method-dataset-collection}
Existing vulnerability datasets do not track the full commit history of vulnerability reintroduction and subsequent fixes. 
To address this, we constructed a ground truth dataset by extending existing datasets that capture ImageMagick vulnerability records.
We first extracted existing ImageMagick vulnerability commits from two exemplar datasets: 
    BigVul~\citep{fan_cc_2020} and 
    DiverseVul~\citep{chen_diversevul_2023}.
For our initial case study, we filtered for commits that modified only one file.

To filter for commits that specifically reintroduced vulnerabilities, we analyzed subsequent commits from the identified set of ImageMagick vulnerability fixing commits. 
To determine whether regressions occurred that reintroduced vulnerabilities, we leveraged the SZZ algorithm, a widely known approach for identifying defect inducing source code~\citep{sliwerski_when_2005}.
For each vulnerability fixing commit, we used SZZ to trace the modified lines of code through the version control history to identify all modifying commits, thereby identifying the proper corrective fix if available. 
Additionally, we also manually evaluated the commit message attached to each SZZ identified corrective commits to determine if the commit resolved a reintroduced vulnerability.

To further refine our dataset and ensure the correctness of each vulnerability reintroducing commit, we leveraged five LLMs: 
    Claude-3.5-Sonnet (Anthropic), 
    GPT-4o (OpenAI), 
    DeepSeek Chat (DeepSeek AI), and 
    Grok-2 (xAI).
Based on our sample-level evaluation, we selected Claude as the primary model for constructing the vulnerability reintroducing dataset. 
All Claude-generated labels were then manually reviewed as a second validation step to ensure accuracy.
Our prompt to do so is presented in Listing 3.

\captionsetup{justification=centering}
\begin{figure}[t]
    \small
    \centering
    \begin{tcolorbox}[
        title=\textbf{Listing 3: Anthropic Claude 3.5 Sonnet System Prompt},
        colback=blue!3!white,        
        colframe=blue!60!black,      
        colbacktitle=blue!60!black,
        coltitle=white,              
        width=0.97\columnwidth, 
        boxrule=0.7pt, 
        arc=0mm, 
        sharp corners,
        left=5pt, right=5pt, top=4pt, bottom=4pt,
        before skip=4pt, after skip=4pt
    ]
    \renewcommand{\arraystretch}{1.1}
    \setlength{\parskip}{2pt}

    \textbf{Research context:} I am conducting research on how software \textbf{security} vulnerabilities evolve over time, 
    specifically focusing on situations where a fix for one \textbf{security} vulnerability unintentionally introduces a new \textbf{security} vulnerability.

    \vspace{4pt}
    \textbf{Task:} You will be provided with details from two commits:
    \begin{itemize}
        \item \textbf{Previous Fix Commit} – A commit that fixed a known vulnerability.
        \item \textbf{Future Candidate Commit} – A later commit that modifies the same or nearby code.
    \end{itemize}
    Your task is to determine whether the \textbf{candidate commit} is fixing a new vulnerability that was introduced by the \textbf{previous fix}.

    \vspace{4pt}
    \textbf{Previous Fix Details}
    \begin{itemize}
        \item \textbf{Commit ID:} \{commit\_hash\}
        \item \textbf{Commit Message:} \{previous\_fix\_message\}
        \item \textbf{Code Changes (Diff Format):} \{previous\_fix\_diff\_content\}
    \end{itemize}

    \textbf{Future Candidate Details}
    \begin{itemize}
        \item \textbf{Commit Message:} \{future\_commit\_message\}
        \item \textbf{Code Changes (Diff Format):} \{future\_diff\_content\}
    \end{itemize}

    \vspace{4pt}
    \textbf{Your Response Format (Strictly Follow This JSON Format)}

    \begin{tcolorbox}[
        colback=white,
        colframe=blue!40!black,
        left=4pt, right=4pt, top=2pt, bottom=2pt,
        width=\linewidth, boxrule=0.4pt,
        arc=0mm, sharp corners
    ]
    \small
\texttt{\{"answer": "Yes" or "No",}\\
\texttt{"reasoning": "Detailed explanation of why the candidate commit is or is not fixing a vulnerability introduced by the previous fix."}
    \end{tcolorbox}

    \textbf{Important condition:} \emph{If the previous fix is incomplete, the answer is "No" since it did not introduce a new vulnerability. 
    Similarly, if the previous fix did not properly fix the issue, the answer is still "No" unless a new vulnerability is created.}
    \end{tcolorbox}

    \caption{
        Anthropic Claude 3.5 Sonnet LLM system prompt used to evaluate commit pairs where one commit fixed a vulnerability and a subsequent commit potentially addressed a new vulnerability introduced by that fix. 
    }
    \label{fig:prompt_single}
    \vspace{-5pt}
\end{figure}

Our dataset collection resulted in 81 confirmed vulnerability reintroducing-fixing commit pairs for ImageMagick from an initial set of 175 and 225 commits from BigVul and DiverseVul respectively.
Of the 81 pairs, 25 originated from BigVul~\citep{fan_cc_2020} and 56 originated from DiverseVul~\citep{chen_diversevul_2023}, however only 76 have associated CVEs. 
Our dataset is publicly available~\citep{shimmi_2026_18251736}

\subsubsection{CVE Selection Criteria}
\begin{table}[]
    \small
    \setlength{\abovecaptionskip}{2pt}
    \setlength{\belowcaptionskip}{0pt}
    \caption{
    CVEs analyzed and their CWE vulnerability classification and CVSS severity score.
    }
    \begin{tabular}{ccc}
        \textbf{CVE} & \textbf{CWE} & \textbf{CVSS} \\ \hline
        CVE-2016-4564
        & CWE-119 & 9.8 \\
        CVE-2017-16546
        & CWE-119 & 8.8 \\
        CVE-2018-11625
    & CWE-125 & 8.8 \\
        CVE-2019-13299
        & CWE-125 & 8.8
    \end{tabular}
    \label{table:tableB}
   
\end{table}
 \vspace{-5pt}
From the 76 candidate pairs of vulnerability reintroducing commits identified in~\Cref{sec:method-dataset-collection}, we randomly selected four pairs.
Our selection criteria were that
    each pair was identified to resolve a known CVE with at least a score of 8.8 while reintroducing a new vulnerability,
    that each vulnerability pair is non-overlapping, and
    each vulnerability had to be identified in a unique year.
Our candidate vulnerabilities are presented in~\Cref{table:tableB}.

\subsubsection{Software Process Metrics Analysis}
For longitudinal computation of process metrics, we utilized the PRIME tool~\citep{synovic_snapshot_2022}.
PRIME can compute 
    bus factor, 
    issue density, and
    issue spoilage
among others.
For bus factor, we computed the bus factor for the entire project every six months.
For issue density and spoilage, we computed the metric per six months, but for each CVE in our case study, we computed the metric per week for 20 weeks prior to, during, and 20 after the reintroduced vulnerability persisted in the project.

\subsection{Data Analysis}
In this section, we discuss
    an overview of the dataset of 76 candidate vulnerability reintroducing-fixing commit pairs identified in~\Cref{sec:method-dataset-collection}, and
    our longitudinal analysis of 
        bus factor,
        issue density, and
        issue spoilage
    as it relates to vulnerability reintroduction activities.

\subsubsection{ImageMagick Vulnerabilities per Year:}
\begin{table}[]
    \small
    \setlength{\abovecaptionskip}{2pt}
    \setlength{\belowcaptionskip}{0pt}
    \caption{
        Breakdown of the number of vulnerability reintroducing commits with identifiable CVEs in ImageMagick from 2015 to 2021.
        }
    \begin{tabular}{ccc}
        \textbf{Year} & \textbf{\# Of Vuln. Commits} & \textbf{Avg. CVSS} \\ \hline
        2015 & 5 & 6.5 \\
        2016 & 12 & 7.4 \\
        2017 & 24 & 7.1 \\
        2018 & 10 & 6.8 \\
        2019 & 11 & 7.7 \\
        2020 & 13 & 5.0 \\
        2021 & 1 & 5.5
    \end{tabular}
    \label{table:tableA}
 \vspace{-5pt}
\end{table}
\Cref{table:tableA} breaks down the 76 identified vulnerability reintroducing commits with CVEs and their average CVSS scores per year.
Since 2017, the number of vulnerabilities that are reintroduced into ImageMagick has decreased.
Furthermore, the average CVSS score of vulnerabilities that reintroduce vulnerabilities has also decreased.
This suggests that ImageMagick’s contributors are becoming increasingly effective at identifying, isolating, and mitigating security flaws prior to including them in the project.
It also indicates a maturation of the project’s secure development practices, where vulnerability resolution attempts are less likely to result in the reintroduction of high-impact security issues.

\subsection{Bus Factor}
\begin{figure}[h]
    \centering
    \setlength{\abovecaptionskip}{2pt}
    \setlength{\belowcaptionskip}{0pt}
    \includegraphics[width=\linewidth]{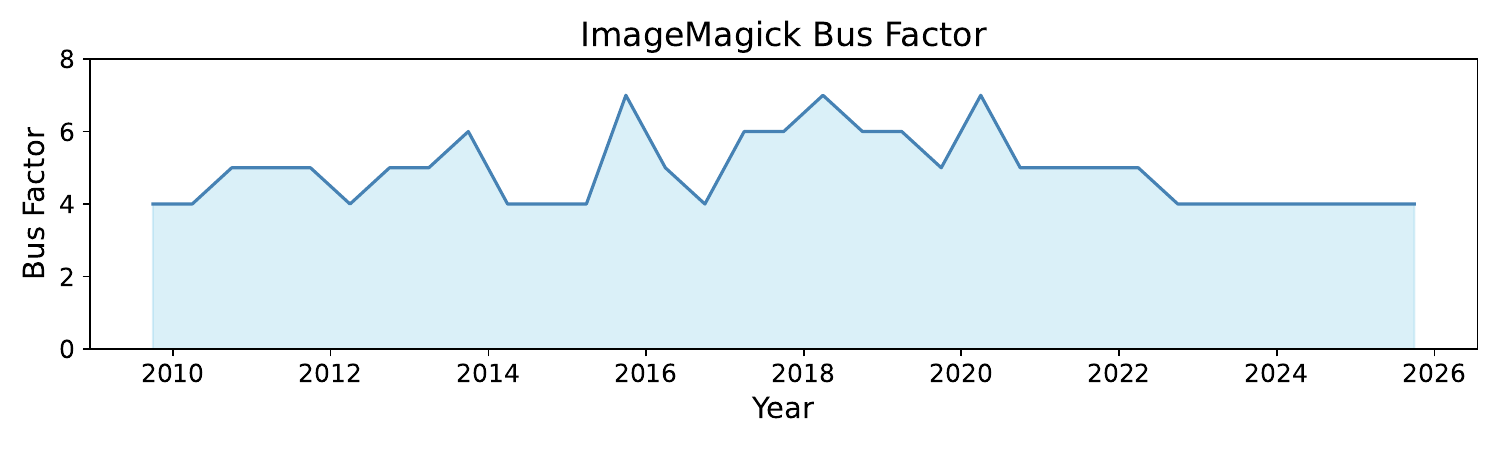}
    \caption{
    ImageMagick maintains a relatively healthy bus factor, with contributions from 4-7 unique maintainers every six months, facilitating distributed handling of vulnerability resolutions.
    }
    \label{fig:figA}
\end{figure}
\vspace{-5pt}

ImageMagick's bus factor per six months metric is presented in~\Cref{fig:figA}.
ImageMagick has at least four---and sometimes as many as seven---contributors during a six-month period.
This guarantees that at any period, there have been a sufficient number of contributors who can introduce and resolve vulnerabilities.

\subsection{Issue Density}
\begin{figure*}[h]
    \centering
    \setlength{\abovecaptionskip}{2pt}
    \setlength{\belowcaptionskip}{0pt}
    \includegraphics[width=0.8\linewidth]{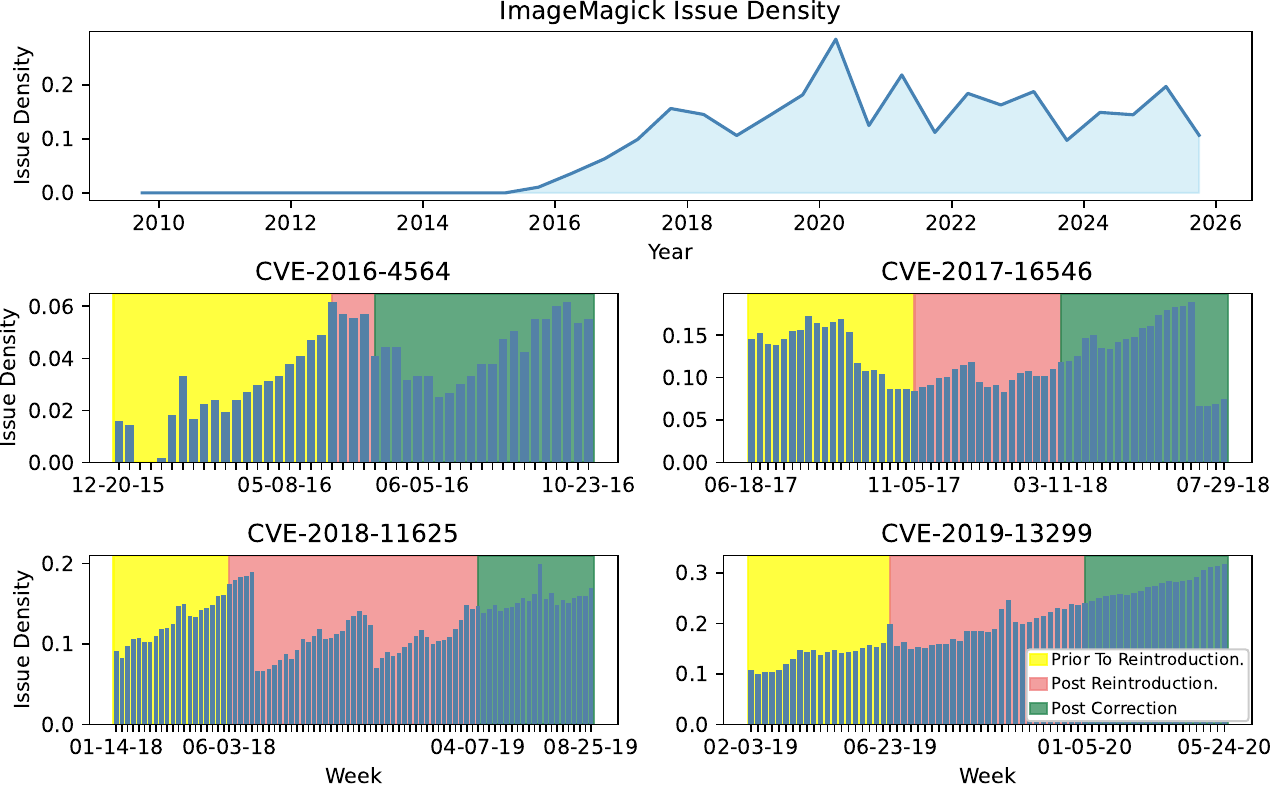}
    \caption{
    Since peaking at 28\% in 2022, issue density has declined to 16.4\%. 
    Fluctuations are tied to specific vulnerabilities: CVE-2016-4564 and CVE-2019-13299 saw density increases driven by rising open-issue rates. 
    Conversely, CVE-2017-16546 and CVE-2018-11625 experienced declines due to code deletion and high issue-closure rates, respectively.
    }
    \label{fig:figB}
\end{figure*}


ImageMagick's and selected CVEs' issue density is presented in~\Cref{fig:figB}.
ImageMagick's issue density per six months is low, at only 28\% of open issues to total project size at its peak in 2020.
Our findings suggest that vulnerability reintroduction is often accompanied by temporary disruptions in issue management efficiency, reflecting the project’s fluctuating capacity to balance bug triage and ongoing development. 

\cveparagraph{CVE-2016-4564}
Prior to the fix for CVE-2016-4564 in which a 
vulnerability was reintroduced, issue density was increasing.
After the vulnerability was reintroduced, issue density decreased until after the corrective fix was applied.
When evaluating the rate at which the project size was changed --- 
    measured in thousands of lines of code per week (i.e., KLOC per week)
--- during this time frame, we see that it experiences little change.
Thus, the issue density of this CVE is reflective of contributors initially closing issues raised by the community, before the community started contributing more issues than the author's could close. 

\cveparagraph{CVE-2017-16546}
Prior to the fix for CVE-2017-16546 in which a 
vulnerability was reintroduced, issue density was decreasing.
After the vulnerability was reintroduced, ImageMagick underwent a wave of increasing and decreasing issue density.
However, after the proper corrective fix was applied, issue density increased.
When evaluating the KLOC per week during this time frame, we see that it initially dips before increasing to it's initial size.
Thus the issue density of this CVE is reflective of both a shrinking code size and closure of open issues. 

\cveparagraph{CVE-2018-11625}
Prior to the fix for CVE-2018-11625 in which a vulnerability was reintroduced, and after the proper corrective fix was applied, ImageMagick experienced increasing issue density.
During the existence of the reintroduced vulnerability, issue density underwent a wave of increasing and decreasing issue density, until issue density continuously rose post the proper corrective fix being applied.
When evaluating the KLOC per week during this time frame, we see that it increases during this time frame, reflecting that the project maintainers were modifying the code and closing issues concurrently during this time frame.

\cveparagraph{CVE-2019-13299}
Prior to the fix for CVE-2019-13299 in which a vulnerability was reintroduced, and after the proper corrective fix was applied, issue density was increasing.
When evaluating the KLOC per week during this time frame, we see that it remains stable.
Thus, the issue density of this CVE is reflective of an increasing amount of issues being opened during this time frame.

\subsection{Issue Spoilage}
\begin{figure*}[h]
    \centering
    \setlength{\abovecaptionskip}{2pt}
    \setlength{\belowcaptionskip}{0pt}
    \includegraphics[width=0.8\linewidth]{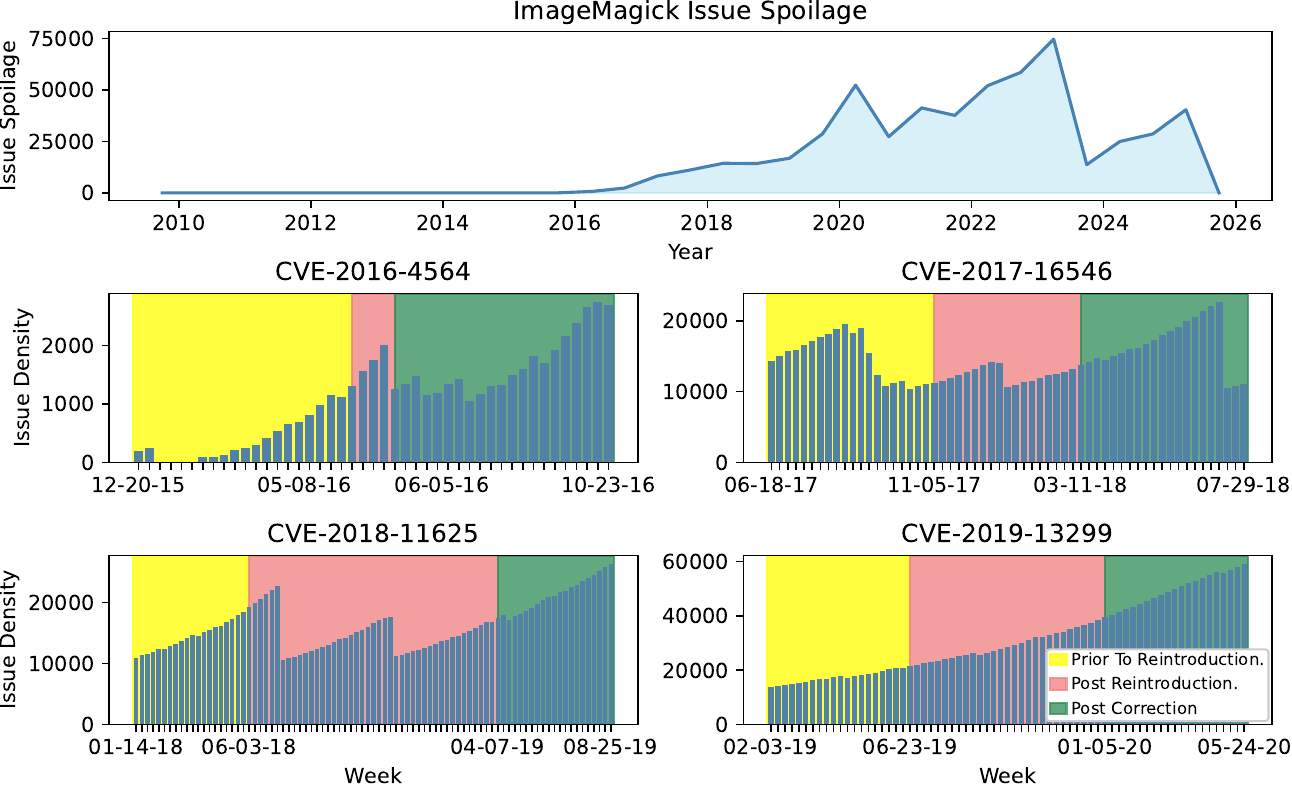}
    \caption{
    After a 2023 peak of 74,657 days, spoilage plummeted to just 25 days. 
    Trends varied by vulnerability: CVE-2016-4564 and CVE-2019-13299 saw rising spoilage following reintroduction—indicating a struggle to resolve issues timely—until corrective patches were applied. 
    In contrast, CVE-2017-16546 and CVE-2018-11625 experienced immediate spoilage dips following vulnerability reintroduction.
    }
    \label{fig:figH}
\end{figure*}

ImageMagick's and selected CVEs' issue spoilage is presented in~\Cref{fig:figH}.
ImageMagick's issue spoilage per six months is high but showing improvement, with over 74,657 days of spoiled issues at its peak in 2023, and currently at 25 days of spoiled issues as of writing.
Our findings indicate that fluctuations in issue spoilage correspond to varying levels of responsiveness and process stability during vulnerability reintroduction periods.

\cveparagraph{CVE-2016-4564}
Prior to the fix for CVE-2016-4564 in which a vulnerability and thereafter, issue spoilage was increasing.
After the proper corrective fix was applied, issue spoilage decreased before increasing again.
This reflects that towards the tail end of the reintroduced vulnerability's existence, project maintainers resolved long open issues before resuming other activities.

\cveparagraph{CVE-2017-16546}
Prior to the fix for CVE-2017-16546 in which a vulnerability was reintroduced, issue spoilage was decreasing. 
After the vulnerability was reintroduced, ImageMagick experienced increasing issue spoilage, a dip, and then increasing issue spoilage after the proper corrective fix was applied.
This reflects that during this reintroduced vulnerabilities life, the maintainers were actively resolving issues while conducting other activities.

\cveparagraph{CVE-2018-11625}
Prior to the fix for CVE-2018-11625 in which a vulnerability was reintroduced and upon reintroduction, issue spoilage was increasing.
While the reintroduced vulnerability was active, ImageMagick experienced increasing issue spoilage, a dip, and then continuous increase after proper resolution.
This reflects maintainers initially closing a significant number of issues, and then continuing to resolve issues throughout the reintroduced vulnerabilities existence.

\cveparagraph{CVE-2019-13299}
Prior to the fix for CVE-2019-13299 in which a vulnerability was reintroduced, and after the proper corrective fix was applied, issue spoilage was increasing. 
This is reflective of the maintainers not being able to close open issues faster than they were being opened.

\section{Threats To Validity}
We report our findings based on an initial study of a subset of vulnerability reintroduction commit pairs from ImageMagick --- a well known image processing project.
The relatively small sample size of analyzed commit pairs limits the generalizability of our conclusions. 
Since ImageMagick is primarily written in C/C++, the results may not generalize to projects in other languages or domains. 
Furthermore, not all open-source projects publicly disclose vulnerabilities to centralized vulnerability databases, potentially limiting the applicability of our approach to the wider open-source community.
To mitigate these threats we relied on existing vulnerability datasets~\citep{fan_cc_2020, chen_diversevul_2023} to identify projects with identified CVE and CWEs to focus on projects of interest to the software engineering community.

Our methodology to identify pairs of vulnerability reintroducing and fixing commits leverages the Anthropic Claude 3.5 Sonnet LLM to validate manually identified pairs from the BigVul vulnerability dataset~\citep{fan_cc_2020}.
Additionally, it autonomously identifies commit pairs from the DiverseVul vulnerability dataset~\citep{chen_diversevul_2023}.
This decision was driven by the significant time and complexity involved in conducting a full manual analysis.

Furthermore, our leveraged bus factor metric~\citep{cosentino_assessing_2015} measures the aggregate number of unique contributors to a project, and does not take into account who contributes to individual directories or files.
Thus, while our findings show that at least four contributors are working on ImageMagick at any one time, it does not enable the evaluation of the engineering activity of each contributor.
To mitigate this, we focus the holistic effort of the team, rather than specific engineers.

\section{Discussion and Future Implications}

This initial case study on ImageMagick demonstrates the value of analyzing software engineering process metrics to understand vulnerability reintroduction patterns. 
Our findings reveal potential insight on how bus factor, issue density, and issue spoilage correlate with vulnerability reintroduction activities. 
These findings highlight several directions for future research.

\myparagraph{Expanding Beyond ImageMagick}
We intend to expand the scope of analysis to multiple open-source systems across diverse domains and programming languages. 
By extending our methodology to diverse projects, we want to assess whether there is any general correlation between software engineering processes and vulnerability reintroduction. 
This large-scale analysis will allow us to determine if certain socio-technical patterns are consistent among projects. 

\myparagraph{Developing Predictive Models for Risky Fixes}
The overarching objective of this research is to enable actionable prediction of vulnerability reintroduction. 
Future work will develop machine learning classifiers trained on longitudinal process metrics, code-level features, and commit metadata to predict vulnerability-fixing commits that are likely to reintroduce new security issues before they are proposed.
Such predictive systems could inform automated regression testing and security-focused quality assurance workflows.

\myparagraph{Investigating Pull Request Dynamics}
Future research will explore how code review processes impact vulnerability reintroduction. 
By analyzing metrics pertaining to pull requests
(e.g.,
    pull request spoilage, 
    review latency, 
    reviewer-to-contributor ratios, 
    comment volume)
we aim to evaluate their relationship with reintroduction risk. 
This analysis will help identify specific pull request practices that contribute to higher vulnerability reintroductions.

\section{Conclusion}

To our knowledge, this is the first work to highlight the critical role of software engineering process metrics in understanding and mitigating the risk of vulnerability reintroduction. 
By shifting focus from file-level vulnerability prediction to analyzing sequences of security fixing commits, we underscore the importance of detecting vulnerability enabling engineering activities longitudinally.
Our approach emphasizes that vulnerability reintroduction is rarely the result of a single isolated action, but instead emerges from cumulative sequences of engineering activities and socio-technical conditions.

\begin{acks}
The work in this paper was partially funded by the Office of Naval Research (ONR) (Grant\#: G2A62826).
\end{acks}

\balance
\bibliographystyle{ACM-Reference-Format}


\end{document}